\documentclass[modern]{aastex631}
\usepackage{multirow}
\usepackage{comment}

\shorttitle{Multi-scale stamp classifier}
\shortauthors{Reyes-Jainaga et al.}

\graphicspath{{./}{figures/}}

\begin{document}

\title{Multi-scale stamps for real-time classification of alert streams}

\newcommand\MAS{Millennium Institute of Astrophysics, Nuncio Monse{\~{n}}or S{\'{o}}tero Sanz 100, Providencia, Santiago, Chile}
\newcommand\CMM{Center for Mathematical Modeling, Universidad de Chile, Beauchef 851, Santiago 8370456, Chile}
\newcommand\IDIA{Data and Artificial Intelligence Initiative (IDIA), Faculty of Physical and Mathematical Sciences, Universidad de Chile, Chile.}
\newcommand\UV{Instituto de F\'{\i}sica y Astronom\'{\i}a, Universidad de Valpara\'{\i}so, Chile}
\newcommand\ESODE{European Southern Observatory, Karl-Schwarzschild-Strasse 2, 85748 Garching bei München, Germany}
\newcommand\DAS{Departamento de Astronom\'{\i}a, Universidad de Chile, Chile}
\newcommand\UDEC{Department of Computer Science, Universidad de Concepci\'{o}n, Concepci\'{o}n, Chile}
\newcommand\UDS{Data Science Unit, Universidad de Concepci\'{o}n, Concepci\'{o}n, Chile}
\newcommand\DIE{Department of Electrical Engineering, Universidad de Chile, Av. Tupper 2007, Santiago 8320000, Chile}
\newcommand\DCCUCH{Computer Science Department (DCC), University of Chile, Chile}
\newcommand\Fintual{Fintual Administradora General de Fondos S.A., Santiago, Chile}
\newcommand\DO{Data Observatory Foundation, Santiago, Chile}

\author[0000-0003-3627-0216]{Ignacio Reyes-Jainaga}
\affiliation{\CMM}
\affiliation{\MAS}

\author[0000-0003-3459-2270]{Francisco F\"orster}
\affiliation{\IDIA}
\affiliation{\MAS}
\affiliation{\CMM}
\affiliation{\DAS}

\author[0000-0002-8722-516X]{Alejandra M. Muñoz Arancibia}
\affiliation{\MAS}
\affiliation{\CMM}

\author[0000-0002-2720-7218]{Guillermo Cabrera-Vives}
\affiliation{\UDEC}
\affiliation{\MAS}
\affiliation{\UDS}

\author[0000-0001-7868-7031]{Amelia Bayo}
\affiliation{\UV}
\affiliation{\ESODE}

\author[0000-0002-8686-8737]{Franz E. Bauer}
\affiliation{Instituto de Astrof\'{\i}sica and Centro de Astroingenier{\'{\i}}a, Facultad de F\'{i}sica, Pontificia Universidad Cat\'{o}lica de Chile, Casilla 306, Santiago 22, Chile.}
\affiliation{\MAS}

\author[0000-0002-2045-7134]{Javier Arredondo}
\affiliation{NeuralWorks, Santiago, Chile}

\author[0000-0003-3455-9358]{Esteban Reyes}
\affiliation{\Fintual}

\author[0000-0003-0006-0188]{Giuliano Pignata}
\affiliation{Instituto de Astrof\'{\i}sica, Facultad de Ciencias Exactas, Universidad Andres Bello, Avda.  Fern\'endez Concha 700, Las Condes, Chile}
\affiliation{\MAS}

\author[0000-0002-0855-1849]{A. M. Mour\~ao}
\affiliation{Departamento de Física and  CENTRA-Center for Astrophysics and Gravitation, Instituto Superior Técnico,  Universidade de Lisboa, Portugal}

\author{Javier Silva-Farf\'an}
\affiliation{\DAS}

\author[0000-0002-1296-6887]{Llu\'is Galbany}
\affiliation{Institute of Space Sciences (ICE, CSIC), Campus UAB, Carrer de Can Magrans, s/n, E-08193 Barcelona, Spain.}
\affiliation{Institut d’Estudis Espacials de Catalunya (IEEC), E-08034 Barcelona, Spain.}

\author[0000-0001-6519-0923]{Alex \'Alvarez}
\affiliation{\CMM}

\author[0000-0001-6172-9362]{Nicol\'as Astorga}
\affiliation{\MAS}
\affiliation{\CMM}
\affiliation{\DIE}

\author[0000-0002-8585-8035]{Pablo Castellanos}
\affiliation{\UDS}

\author[0009-0004-2791-8601]{Pedro Gallardo}
\affiliation{\MAS}
\affiliation{\UDS}

\author[0000-0002-7003-5087]{Alberto Moya}
\affiliation{\MAS}
\affiliation{\UDS}
\affiliation{\DCCUCH}

\author[0000-0002-6984-2965]{Diego Rodr\'iguez}
\affiliation{\DO}

\begin{abstract}

In recent years, automatic classifiers of image cutouts (also called ``stamps'') have shown to be key for fast supernova discovery. The Vera C. Rubin Observatory will distribute about ten million alerts with their respective stamps each night, enabling the discovery of approximately one million supernovae each year. A growing source of confusion for these classifiers is the presence of satellite glints, sequences of point-like sources produced by rotating satellites or debris. The currently planned Rubin stamps will have a size smaller than the typical separation between these point sources. Thus, a larger field of view stamp could enable the automatic identification of these sources. However, the distribution of larger stamps would be limited by network bandwidth restrictions. We evaluate the impact of using image stamps of different angular sizes and resolutions for the fast classification of events (AGNs, asteroids, bogus, satellites, SNe, and variable stars), using data from the Zwicky Transient Facility. We compare four scenarios: three with the same number of pixels (small field of view with high resolution, large field of view with low resolution, and a multi-scale proposal) and a scenario with the full stamp that has a larger field of view and higher resolution. Compared to small field of view stamps, our multi-scale strategy reduces misclassifications of satellites as asteroids or supernovae, performing on par with high-resolution stamps that are 15 times heavier. We encourage Rubin and its Science Collaborations to consider the benefits of implementing multi-scale stamps as a possible update to the alert specification.

\end{abstract}


\section{Introduction} \label{sec:intro}

The advent of new large etendue survey telescopes, such as the Vera C. Rubin Observatory \citep{Ivezic_LSST_2019}, will revolutionize time-domain astronomy by discovering changes in the sky at a higher rate than ever before. Surveys such as the Zwicky Transient Facility (ZTF, \citealt{bellm2018zwicky}) and the Asteroid Terrestrial-impact Last Alert System (ATLAS) \citep{tonry2018atlas} are already producing hundreds of thousands of alerts per night where the brightness or location of a source change \citep{Patterson_Zwicky_2018, Heinze_First_2018}. This number is expected to grow to millions when the Rubin Observatory's Legacy Survey of Space and Time (LSST) starts operating \citep{Ivezic_LSST_2019, Ridgway2014variable}.  

\begin{figure}[htbp]
    \centering
    \includegraphics[width=0.95\textwidth]{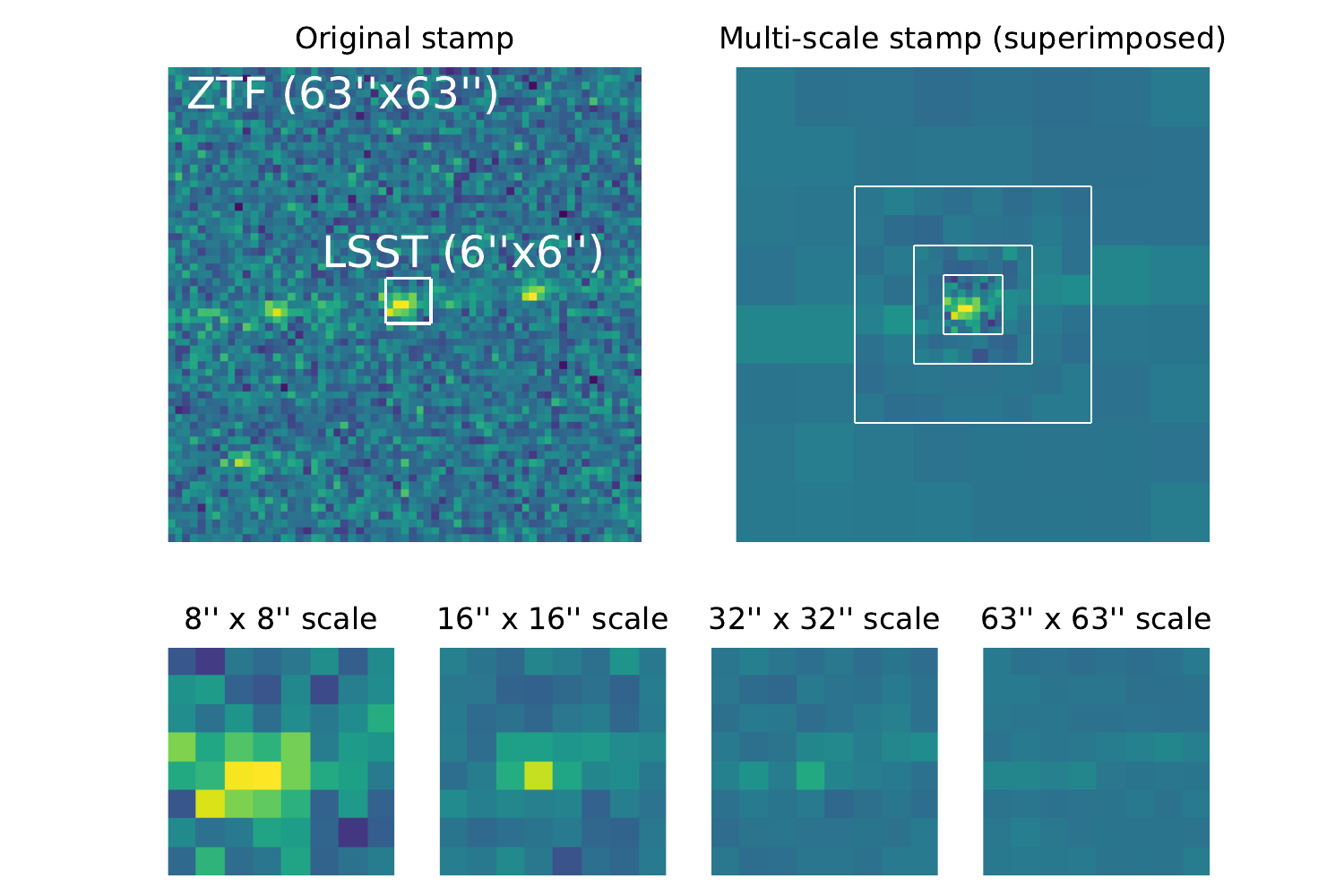}
    \caption{Visualization of the proposed multi-scale stamps. The original stamp shows a stamp from an alert of ZTF, and the small square in the middle represents the size of an LSST stamp. The multi-scale stamp on the upper right side shows the same stamp from ZTF, but represented as four images of 8 by 8 pixels of different scales. The four images are presented in the lower part of the figure with their corresponding Field of View.}
    \label{fig:multiscale_visualization}
\end{figure}

To facilitate the use of such large volumes of data to the scientific community, astronomical broker systems are being developed. Brokers are complex systems that ingest, process, and serve astronomical alert streams. Processed alerts delivered by the brokers to the community and other systems usually include cross-match associations with other catalogs, prioritization for follow-up observations, and classification of the sources. Full-stream community brokers for LSST include ALeRCE \citep{Forster_ALeRCE_2021}, AMPEL \citep{nordin2019transient}, ANTARES \citep{narayan2018machine}, Babamul, Fink \citep{moller2021fink}, Lasair \citep{smith2019lasair}, and Pitt-Google\footnote{\url{https://pitt-broker.readthedocs.io/en/latest/}}.

One of the most important tasks performed by brokers is the automatic classification of sources triggering alerts. First approaches for automatic classification of astronomical images used classical machine learning methods based on handmade features calculated over the candidate stamps \citep{romano2006supernova, bailey2007find, bloom2012automating, brink2013using, goldstein2015automated, forster2016high}. Newer methods like Convolutional Neural Networks \citep[ConvNets or CNNs;][]{fukushima1980self, lecun1998gradient} have outperformed classical machine learning approaches by directly learning features from the data. They were first applied in astronomy to the prediction of galaxy morphologies \citep[e.g.][]{dieleman2015rotation, gravet2015catalog, perez2019multiband, barchi2020machine, cheng2021galaxy, walmsley2022galaxy} and since then they have been applied to other time-domain survey challenges such as real/bogus alert classification \citep{ cabrera2016supernovae, cabrera-vives_deep_2017, reyes2018enhanced, duev2019real, turpin2020vetting, yin2021supernovae}, time domain classification \citep{carrasco2019deep, aguirre2019deep, gomez2020classifying}, and transient host identification \citep{forster2022delight}, among others.

Performing alert classification in a fast and accurate way is particularly important for ALeRCE, which is currently doing same-night classification based on single alert image stamps from the ZTF public alert stream \citep{carrasco2021alert}. This model uses the science, template, and difference images of only the first detection as inputs to classify alerts into five classes: active galactic nuclei (AGN), supernovae (SNe), variable stars (VS), asteroids, and bogus. \cite{carrasco2021alert} highlights that $\sim 70\%$ of the candidates sent to the Transient Name Server\footnote{\url{https://www.wis-tns.org/}} (TNS) were reported within one day after the detection. To date, ALeRCE has reported over 17,000 supernova candidates to the TNS based on its stamp classifier\footnote{\url{https://www.wis-tns.org/stats-maps}}. During the 2021-2022 period, 4495 supernovae were spectroscopically confirmed in TNS, and ALeRCE sent the first discovery report for 1273 of them ($\sim 28.32 \%$). Overall, the use of cutout stamps has proven essential for early supernova detection by ALeRCE, allowing the discovery of promising transient candidates with just one observation.

Alerts from survey telescopes, such as ZTF or LSST, include image stamps centered at the coordinates of the emitted alert. The planned uncompressed alert image stamps for LSST will contribute approximately 20\% of the total alert packet size and will be at least $30\times30$ pixels in size \citep[DMTN-102,][]{DMTN-102}. The size of the stamps that the Rubin Observatory will distribute to its alert brokers is mostly limited by the required network bandwidth, given the large number of alerts that are expected in every exposure ($\sim$ 40000 alerts per exposure on average when observing close to the Milky Way). Assuming a pixel size of $0.2\arcsec$, the field of view (FoV) of each stamp would be approximately $6\arcsec\times6\arcsec$. This poses problems for classifying objects when its relevant context extends outside this FoV. This would be the case for many SNe, where for nearby ones the host galaxy could not fall fully in the FoV.

In this work, we propose a multi-scale image stamp approach for the rapid classification of astronomical alerts as shown in Figure \ref{fig:multiscale_visualization}, that would aid the early discovery of astronomical variable sources, offering a trade-off between a larger FoV and alert size. We train a ConvNet over multi-scale images, and evaluate our strategy using real ZTF alerts. Our training set includes bogus alerts, asteroids, SNe, variable stars, AGN, and satellites. The latter was added to the ALeRCE taxonomy given the relevance that they have recently shown for contaminating the night--sky \citep{karpov2022impact}. Our work uses a similar methodology to DELIGHT \citep{forster2022delight}, where the use of multiresolution images is proposed to identify the host galaxies of extragalactic transients. DELIGHT is able to correctly identify host galaxies using 32 KB multiresolution images, where the largest FoV (and lower resolution) image was $120\arcsec\times 120\arcsec$. Here, we use a maximum FoV of $63\arcsec\times 63\arcsec$, which is the FoV of the images in the ZTF alert stream \citep{masci_zwicky_2018}, and compare different strategies such as using the full images, a cropped version of them, low resolution images of the whole FoV, and our proposed multi-scale approach. We will refer to these strategies as ``Full'', ``Cropped'', ``Low resolution'' or ``low\_res'', and ``Multi-scale'', respectively.

The outline of this paper is as follows. In section \ref{sec:experimental_setting} we describe the experimental setting: the data used, the multi-scale structure proposed for the stamps and other scenarios used for comparison, and the classifiers used. In section \ref{sec:results} we present the metrics and matrices for all the evaluated scenarios and some analysis over additional unlabeled data. In section \ref{sec:discussion} we compare the results of the different scenarios and propose a multi-scale strategy for the LSST stream image cutouts. In section \ref{sec:conclusions} we present the conclusions of this work.

\section{Experimental setting} \label{sec:experimental_setting}

\subsection{Data}

The data used comes from the ZTF public alert stream, described in \citet{masci_zwicky_2018}, which is processed and distributed by the ALeRCE broker \citep{Forster_ALeRCE_2021}. An alert is triggered by ZTF each time there is a 5-sigma detection in a difference image (the science image minus the template image). A single astronomical object can cause multiple alerts to be generated if this 5-sigma threshold is surpassed in different observations. For example, a supernova will cause alerts to be generated for every observation in which it is detected with at least 5-sigma in the difference image.

Each alert contains information about the event such as magnitude, time, position, and other metadata, and also an image cutout of the observation. Three cutouts or ``stamps'' are contained in each alert: science, template, and difference. The science stamp comes from the current observation, the template is a historic reference (usually an image stack) of how that portion of the sky looks, and the difference is a subtraction\footnote{The images are convolved before substraction to match their PSF.} of the science and template stamps. The stamps are $63\times63$ pixels in size with a pixel size of $1\arcsec$. In rare cases, if an alert occurs at the edge of the sensor, it will create a non-squared stamp. For simplicity, non-squared stamps were discarded in this work, although they could be used in future work if they are ``repaired'' with the appropriate padding to make them squared. 

The dataset used in this work is an extension of the one from \citet{carrasco2021alert}, which includes AGN, variable stars, asteroids, SNe and bogus classes. Please check \citet{carrasco2021alert} to see some stamp examples of each class. In this work, we incorporate more examples of the mentioned classes as well as a new ``satellite'' class. These ``satellite'' events were identified by the ALeRCE broker team through visual inspection of the ZTF alerts while looking for supernova candidates using the Supernova Hunter\footnote{\url{https://snhunter.alerce.online}}, and have the same characteristic shape as the objects presented by \citet{karpov2022impact}. These satellite glints often look like many individual sources evenly spaced in a straight line, resembling a pearl necklace. 

The number of AGN, asteroids, bogus, satellites, SNe and variable stars in the labeled set are 9,774; 9,180; 15,950; 580; 3,615; and 10,211; respectively. The additional samples with respect to \citet{carrasco2021alert} came from cross matching recent ZTF alerts against confirmed SNe available on TNS, as well as new identified bogus events by the ALeRCE team. Only one alert chosen randomly was used per each ZTF object in the dataset to avoid information leaks at the moment of partitioning the data for training and evaluation. 

We apply some cuts to ensure that the labeled set is pure. By visually inspecting the bogus dataset, 48 stamps previously tagged as bogus were converted to satellite because they have clear satellite glint shapes. Any alerts labeled as bogus but having a cross-match with a known SN, variable star or AGN, were removed from the labeled set to avoid possible mislabeling. Additionally, all alerts labeled as bogus associated with a single ZTF object with 10 or more detections were removed because they could be real astrophysical events. A sample of 12 bogus ZTF objects with more than 10 detections were manually confirmed as bogus and included to the bogus dataset. An example of this kind of object are very bright stars that saturate the detector and can trigger bogus alerts in many different epochs.

For the asteroid class, only sources with a single detection were kept, as in principle an asteroid should not have more than one detection in a given position of the sky. In \citet{carrasco2021alert}, a ZTF object was labeled as asteroid if one of its alerts has a cross-match with the MPC. The variable stars dataset was built from the ALeRCE dataset \citep{sanchez-saez_alert_2021} taking random samples from Eclipsing binaries (2339 samples), Long Period Variables (1167 samples), RR Lyrae stars (1158 samples), RS Canum Venaticorum stars (1171 samples), Young Stellar Objects (1168 samples), Delta Scuti stars (1175 samples), Cepheid stars (973 samples), ZZ Ceti stars (9 samples) and other types of periodic sources (1051 samples). A similar procedure was followed for the AGN class, composed of AGN (3916 samples), Blazar (1932 samples) and Quasi-stellar objects (3926 samples).

The classification models trained and evaluated in this paper receive three kinds of stamps as input: science, template and difference. They also receive the position in the sky of the alert. The RA/Dec coordinates were mapped to the surface of a unitary sphere, so the models see a 3-dimensional representation of the positions (i.e. instead of representing the positions as two angles, they are expressed as the x, y, z coordinates of the points in the surface of the sphere). In this representation the Galactic, Ecliptic and Equatorial planes are in fact planes in three dimensions, which is not the case for 2-dimensional representations (as RA/Dec) where some planes appear as curves. We expect this 3D representation will help the model to learn the classification task more easily.

The stamps were preprocessed by replacing NaN values, infinite values and numbers larger than $10^{10}$ (in absolute value) with zeros. To preserve that information, a binary mask that indicates the position of previously invalid values was given to the models. For each stamp we compute the absolute values and then we take the minimum and the percentile 99. This minimum value is subtracted and then the stamp values are divided by the difference between the 99 percentile and the minimum (plus $10^{-8}$ in the denominator to avoid dividing by zero). This normalization keeps the zero values at zero, but scales the stamps to make their values more comparable. As a last step, all the values are clipped to the range between -2 and 2, because any extreme value can disrupt the training process of the models.

The dataset was split into training, validation and testing in a 60\%, 20\% and 20\% proportion, respectively. The splitting was stratified, so that the proportions of the 6 classes are the same between sets.

\subsection{Stamp scale comparison}

To evaluate the effect of using multiple scales and compare our strategy with other approaches, we tested four different scenarios.

\begin{enumerate}
    \item ``Full'' stamps: $63\times63$ pixel stamps (FoV of $63\arcsec\times63\arcsec$). Size of 3969 floating point numbers. This is the full size of the stamps sent by ZTF in their public alerts.
    
    \item ``Cropped'' stamps: $16\times16$ pixel stamps (FoV of $16\arcsec\times16\arcsec$). Size of 256 floating point numbers. This scenario evaluates the impact of using a smaller FoV by cropping the stamps.
    
    \item ``Low resolution'' stamps: $16\times16$ pixel stamps (FoV of $63\arcsec \times63\arcsec$). Size of 256 floating point numbers. It has the same size (in terms of floating point numbers) as the ``cropped'' scenario, but it sacrifices resolution to gain a larger FoV.

    \item ``Multi-scale'' stamps: 4 multi-scale stamps with a total size of 256 floating point numbers:
        \begin{itemize}
            \item $8\times8$ pixels, with a $1\arcsec$ pixel width. FoV of $8\arcsec \times 8\arcsec$.
            \item $8\times8$ pixels, with a $2\arcsec$ pixel width. FoV of $16\arcsec \times 16\arcsec$.
            \item $8\times8$ pixels, with a $4\arcsec$ pixel width. FoV of $32\arcsec \times 32\arcsec$.
            \item $8\times8$ pixels, with an $8\arcsec$ pixel width. FoV of $63\arcsec \times 63\arcsec$ (ZTF has a $63\times63$ stamp size).
        \end{itemize}
        These stamps have the same number of floating point numbers as the ``cropped'' and ``low\_res'' scenarios. Figure \ref{fig:multiscale_visualization} shows an example of a multi-scale stamp of the class ``satellite''.

\end{enumerate}

\subsection{Classification models}

The classification is done by using ConvNets, following our previous experiences in astronomical stamp classification \citep{carrasco2021alert, reyes_enhanced_2018, cabrera-vives_deep_2017}. The used models are described in the Appendix \ref{sec:appendix:classification_models}.

\section{Results} \label{sec:results}

\subsection{Testing over labeled dataset}

\begin{figure}[htbp]
    \centering
    \includegraphics{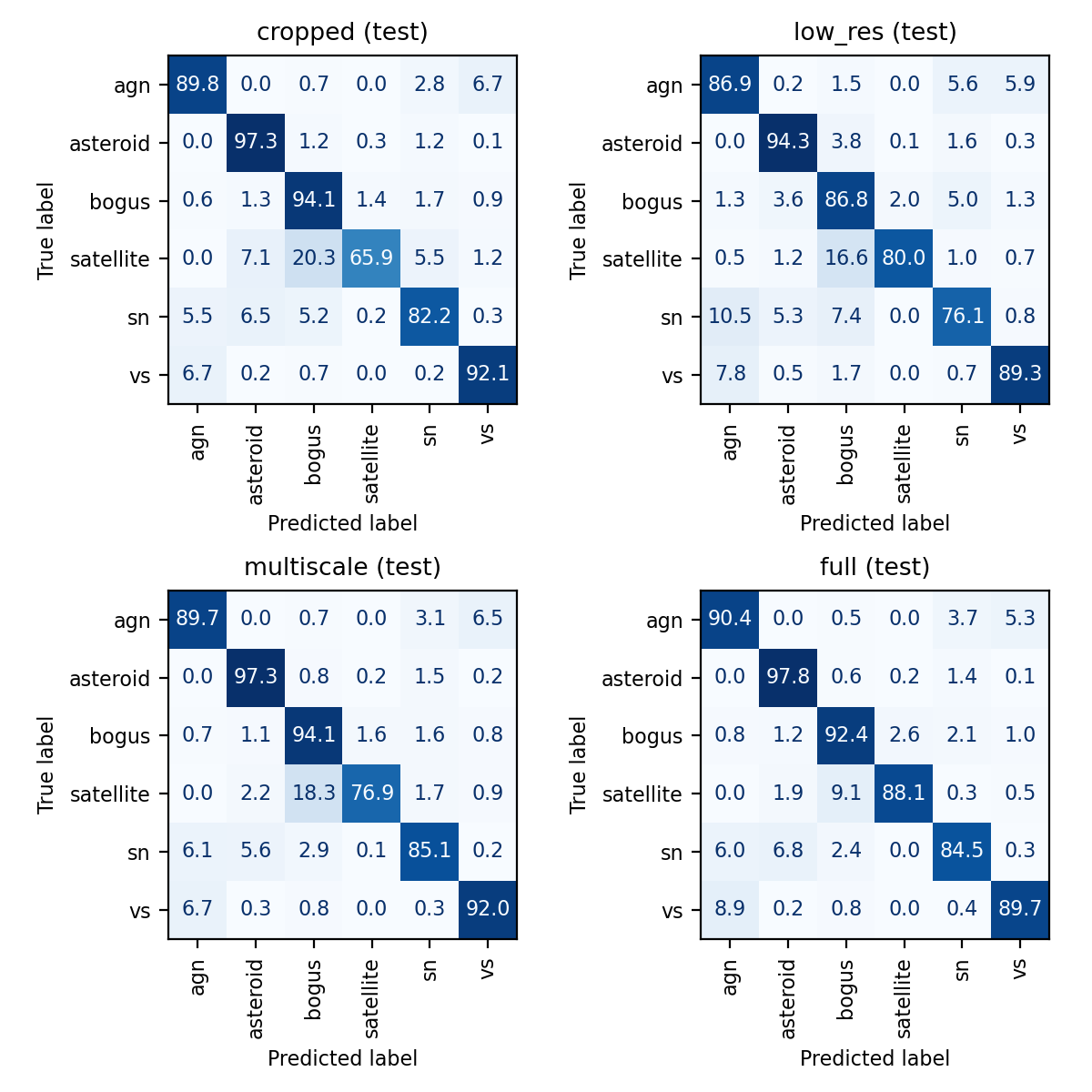}
    \caption{Confusion matrices for the four scenarios, all of them evaluated over the same test set. Each row is normalized to sum 100. AGN means Active Galactic Nuclei, SN means Supernovae, and VS means Variable Stars.}
    \label{fig:confusion_matrices}
\end{figure}

Figure \ref{fig:confusion_matrices} shows the confusion matrices in the four scenarios. Each matrix is the average of 5 network initializations and each row is normalized to sum to 100. The same experiment is also shown in Table \ref{table:metrics_results} using the metrics of f1-score, precision and recall (macro averages). All these metrics are defined in the Appendix \ref{sec:classification_metrics}. This table presents the results for the validation and test set, as well as means and standard deviations for each metric. As the dataset was built by taking a random stamp from a given object and not necessarily the first one, we also retrieved the first available stamp for each object in the test set and evaluated the classification metrics there. This evaluation is closer to what we expect to see in real life, considering that classification of the first alert will have the greatest scientific impact.

The f1-score results of the multi-scale and full-size stamps are statistically similar between each other and higher than the other strategies, despite the multi-scale version weighing approximately 1/15th of the full-size stamps. The next best results are for the cropped stamps model, and finally the low-resolution stamps give the worst f1-score results. Looking at the confusion matrices, the multi-scale results are very similar to the full-sized stamps. One difference is that with the multi-scale stamps more satellites are misclassified as bogus (about two times the mistakes compared to the full-scale scenario), although the bogus alerts misclassified as satellites are less in the multi-scale than the full-scale scenario ($\sim 38 \%$ less mistakes).

When comparing the cropped stamps to the multi-scale approach, one important difference is that 5.5 \% of the satellites are classified as SNe in the cropped scenario, but in the multi-scale this number falls to 1.7 \%. The multi-scale approach also shows a higher recall for the SN class (85.1 \%) than the cropped approach (82.2 \%), highlighting that using cropped stamps with a small FoV has a negative impact in the detection of SNe in comparison with the multi-scale stamp strategy.

\begin{table}[htbp]
\hspace*{-3\leftmargin}
\resizebox{1.2\textwidth}{!}{
\begin{tabular}{c|ccc|ccc|ccc|}
\cline{2-10}
\multicolumn{1}{l|}{}                              & \multicolumn{3}{c|}{Validation set}                                   & \multicolumn{3}{c|}{Test set}                                         & \multicolumn{3}{c|}{Test set (first stamp)}                           \\ \cline{2-10} 
\multicolumn{1}{l|}{}                              & \multicolumn{1}{c|}{F1}     & \multicolumn{1}{c|}{Precision} & Recall & \multicolumn{1}{c|}{F1}     & \multicolumn{1}{c|}{Precision} & Recall & \multicolumn{1}{c|}{F1}     & \multicolumn{1}{c|}{Precision} & Recall \\ \hline
\multicolumn{1}{|c|}{\multirow{2}{*}{Full}}        & \multicolumn{1}{c|}{86.98}  & \multicolumn{1}{c|}{84.91}     & 89.90  & \multicolumn{1}{c|}{86.68}  & \multicolumn{1}{c|}{84.34}     & 90.48  & \multicolumn{1}{c|}{86.65}  & \multicolumn{1}{c|}{84.36}     & 90.43  \\
\multicolumn{1}{|c|}{}                             & \multicolumn{1}{c|}{(0.51)} & \multicolumn{1}{c|}{(0.57)}    & (0.67) & \multicolumn{1}{c|}{(0.42)} & \multicolumn{1}{c|}{(0.48)}    & (0.40) & \multicolumn{1}{c|}{(0.39)} & \multicolumn{1}{c|}{(0.56)}    & (0.35) \\ \hline
\multicolumn{1}{|c|}{\multirow{2}{*}{Cropped}}     & \multicolumn{1}{c|}{85.05}  & \multicolumn{1}{c|}{85.50}     & 84.75  & \multicolumn{1}{c|}{86.19}  & \multicolumn{1}{c|}{85.62}     & 86.88  & \multicolumn{1}{c|}{86.10}  & \multicolumn{1}{c|}{85.58}     & 86.79  \\
\multicolumn{1}{|c|}{}                             & \multicolumn{1}{c|}{(0.41)} & \multicolumn{1}{c|}{(0.74)}    & (0.65) & \multicolumn{1}{c|}{(0.42)} & \multicolumn{1}{c|}{(0.83)}    & (0.51) & \multicolumn{1}{c|}{(0.67)} & \multicolumn{1}{c|}{(1.01)}    & (0.71) \\ \hline
\multicolumn{1}{|c|}{\multirow{2}{*}{Low res}}     & \multicolumn{1}{c|}{82.29}  & \multicolumn{1}{c|}{80.89}     & 84.18  & \multicolumn{1}{c|}{82.69}  & \multicolumn{1}{c|}{80.70}     & 85.55  & \multicolumn{1}{c|}{82.65}  & \multicolumn{1}{c|}{80.61}     & 85.63  \\
\multicolumn{1}{|c|}{}                             & \multicolumn{1}{c|}{(0.21)} & \multicolumn{1}{c|}{(0.41)}    & (0.49) & \multicolumn{1}{c|}{(0.40)} & \multicolumn{1}{c|}{(0.71)}    & (0.30) & \multicolumn{1}{c|}{(0.46)} & \multicolumn{1}{c|}{(0.76)}    & (0.20) \\ \hline
\multicolumn{1}{|c|}{\multirow{2}{*}{multi-scale}} & \multicolumn{1}{c|}{87.20}  & \multicolumn{1}{c|}{86.07}     & 88.54  & \multicolumn{1}{c|}{87.39}  & \multicolumn{1}{c|}{85.99}     & 89.18  & \multicolumn{1}{c|}{87.30}  & \multicolumn{1}{c|}{85.92}     & 89.09  \\
\multicolumn{1}{|c|}{}                             & \multicolumn{1}{c|}{(0.22)} & \multicolumn{1}{c|}{(0.32)}    & (0.30) & \multicolumn{1}{c|}{(0.25)} & \multicolumn{1}{c|}{(0.43)}    & (0.37) & \multicolumn{1}{c|}{(0.34)} & \multicolumn{1}{c|}{(0.55)}    & (0.23) \\ \hline
\end{tabular}}
\caption{Classification metrics for the different scenarios. The values in parenthesis indicate the standard deviation between 5 values. F1, precision, and recall take values between zero and one hundred.}
\label{table:metrics_results}
\end{table}

\subsection{Testing over unlabeled data}

To check how the multi-scale model performs over unlabeled and unseen data, we selected 136,398 objects discovered over a one-year window starting on August 1st, 2020. For each day in the one-year window, we selected all the objects whose first detection happened on that day and then took 500 objects randomly. If less than 500 objects were discovered on a given day, we selected all of them. After that, we checked that no one of these objects were contained in the labeled dataset. Finally, some of the selected objects did not have available stamps or they were not squared stamps, so they could not be used. 49,312 objects were classified as bogus, 54,160 were classified as asteroids, 36,612 were classified as variable stars, 2,338 were classified as AGNs, 1441 were classified as SNe, and 843 were classified as satellites.

\begin{figure}[htbp]
    \centering
    \includegraphics[width=\textwidth]{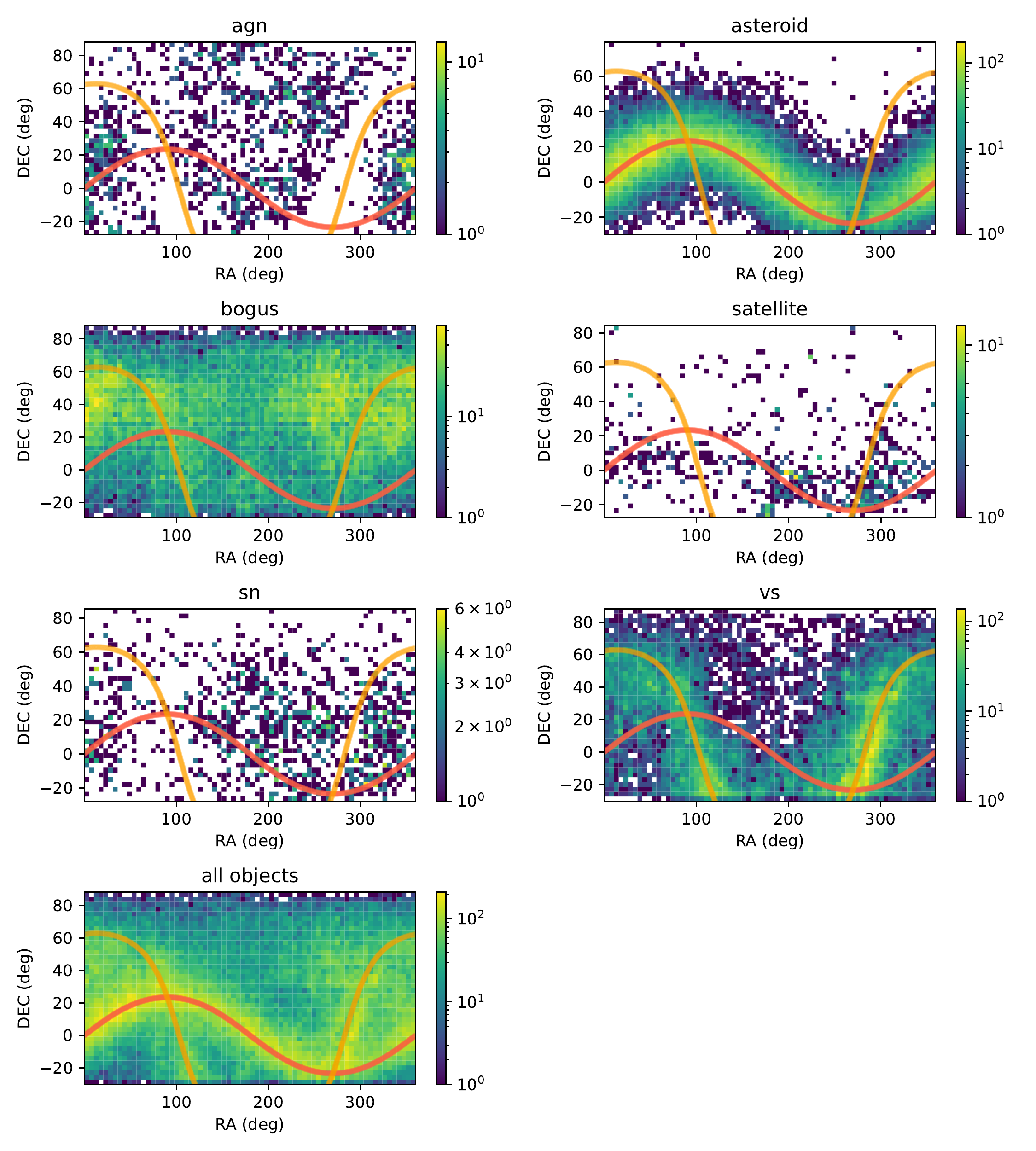}
    \caption{Spatial distribution of a set of unlabeled sources. The sources are separated according to the classification given by the multi-scale stamp classifier. The yellow/orange line corresponds to the Galactic plane and the red line corresponds to the Ecliptic plane.}
    \label{fig:unlabeled_spatial_distribution}
\end{figure}

Figure \ref{fig:unlabeled_spatial_distribution} shows the spatial distribution in the sky of these objects, separated according to the classification given by the multi-scale model. The yellow/orange line corresponds to the Galactic plane and the red line corresponds to the Ecliptic plane. Each class has its expected spatial distribution. AGNs should be everywhere in the sky, except for the occlusion produced by the Milky Way. Asteroids are mainly located in the ecliptic plane. Bogus detections can be found everywhere, but preferentially where more observations are located. Geostationary satellites are flying over the Equator, so they will be to the south given that the observatory is in the northern hemisphere. Satellites with a low orbital inclination (moving close to the Equatorial plane) will also appear to the south for the same reason. SNe and AGNs should be distributed mainly outside of the Galactic plane due to higher extinction in the plane. Variable stars should be close to the Galactic plane, due to their preferential association with young star formation regions, which mostly concentrate in the plane. The last row of Figure \ref{fig:unlabeled_spatial_distribution} shows the distribution of the whole sample.

We can see that the candidates for each class follow approximately the expected spatial distribution. Some classes are harder to check because there are not many candidates found by the classifier for the given sample size. Some voids can be explained by observation biases from the survey and our sample. One obvious case is that ZTF cannot look very far into the southern sky because of its northern hemisphere location. In the Appendix \ref{sec:spatial_distribution_altaz} we show the spatial distribution of the same sources but using the horizontal coordinate system (Altitude - Azimuth).

\begin{figure}[htbp]
    \centering
    \includegraphics[width=\textwidth]{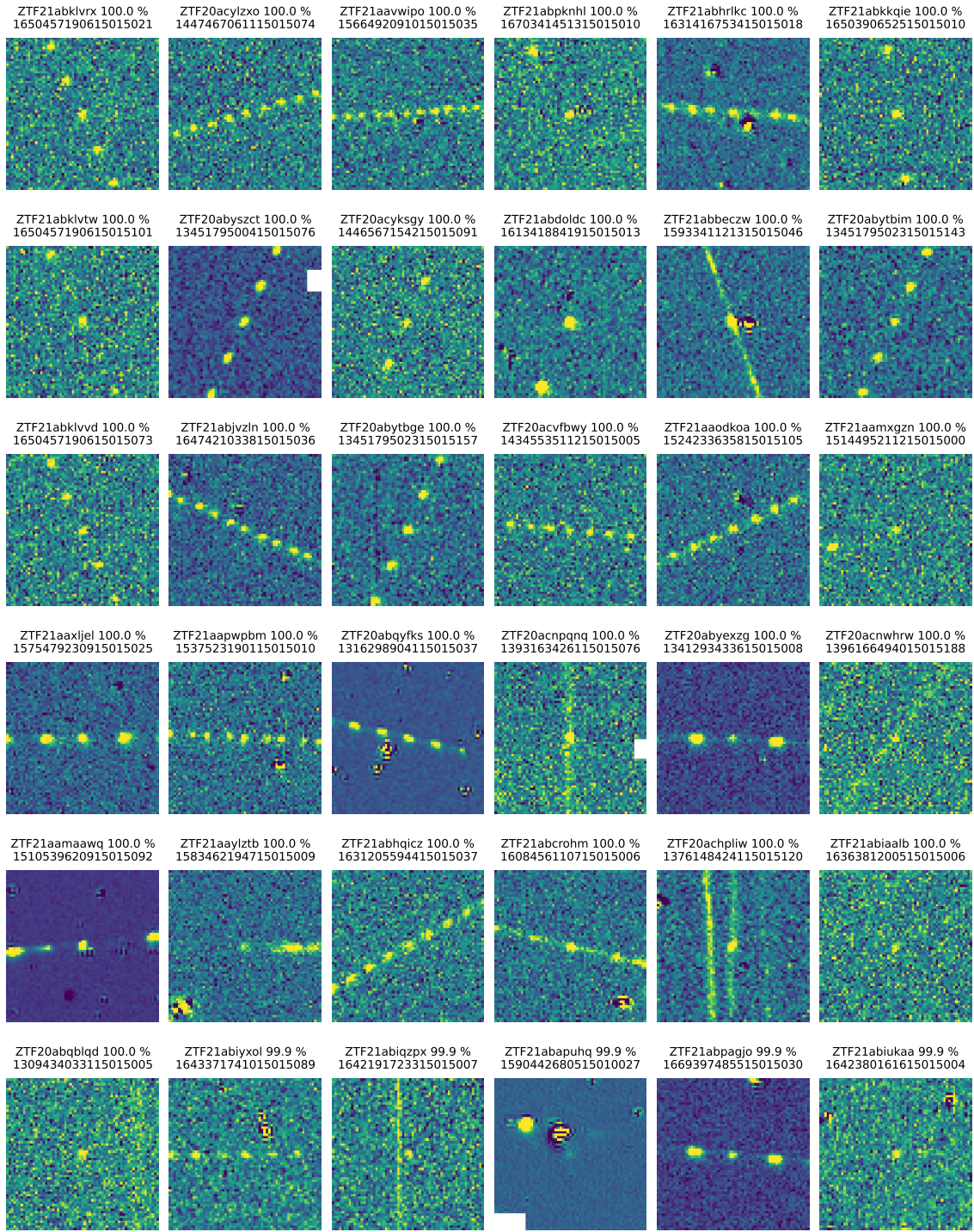}
    \caption{Alerts classified as satellite by the multi-scale stamp classifier. This is a selection of the alerts classified as satellite with the highest confidence. For each alert we show its difference image, object id, and the probability of being a satellite according to the model.}
    \label{fig:satellite_candidates}
\end{figure}

Figure \ref{fig:satellite_candidates} shows 36 stamps with the highest satellite probability from the random sample. Many of them have the characteristic ``pearl necklace'' look of many bright sources evenly spaced in a straight line. 
In addition, we have identified the presence of straight lines across the stamps. Through direct communication with the ZTF team, it has been confirmed that these lines correspond to ``bogus'' asteroids, which we intend to consider as such in the future.

In the Appendix \ref{sec:large_fov_satellite} we show a $500\arcsec\times500\arcsec$ zoom out version for two of these stamps. One of them has a clear ``pearl necklace'' shape, but the other has many streaks across the image. Based on a private communication with the ZTF team, those artifacts are apparently due to readout and clock errors when performing the observation; we aim to include them as a separate class in a future version of our stamp classifier. 

\section{Discussion} \label{sec:discussion}
\subsection{Results analysis}
Looking at the confusion matrices (Figure \ref{fig:confusion_matrices}) and the classification metrics (Table \ref{table:metrics_results}), the multi-scale scenario offers the highest f1-score in the test set, performing on par with the full-size stamp scenario. On the other hand, the low resolution scenario has the worst performance.

The confusion matrices show that the multi-scale model is as good as using the full-size stamps, with the exception of the satellite class, where the recall is lower (76.9 \% in multi-scale vs 88.1 \% in full-size). When comparing the multi-scale proposal with stamps with a small FoV (cropped scenario), the recall in satellite detection improves from 65.9 \% in the cropped scenario to 76.9 \% in the multi-scale strategy. Even more important, a lower fraction of satellites (1.7 \% vs 5.5 \%) is confused with SN in the multi-scale scenario, and that the supernova recall is higher (85.1 \% vs 82.2 \%). This shows that using the multi-scale strategy improves transient detection, compared to using stamps with a very small FoV.

In general, the strategy choice presents a trade-off where depending on the kind of stamp used, one will have different types of errors between the classes. For example, the low resolution model has a higher satellite recall compared with the multi-scale model (80 \% vs 76.9 \%), but the SN recall is higher in the multi-scale scenario (85.1 \% in multi-scale vs 76.1 \% in low resolution).

For AGNs, all the models except the low-resolution have a similar performance, with recall metrics between 89.7 \% and 90.4 \% (86.9 \% for the low-resolution). Asteroid detection follows a similar pattern, with asteroid recall between 97.3 \% and 97.8 \% for all models except low-resolution (94.3 \% for low-resolution). The highest bogus recall is obtained with the multi-scale and the cropped stamps (94.1 \%), but the multi-scale stamps have less satellites and SNe misclassified as bogus. 

The best model for satellite detection is the full-size stamp model, with a satellite recall of 88.1 \%. It is worth highlighting that artificial satellites are not astrophysical objects, so the importance of their detection is primarily to avoid contamination in other astrophysical classes. Given that, the fraction of satellites misclassified as astrophysical objects goes from 13.8 \% in the cropped scenario with a small FoV, to 4.8 \% in the multi-scale proposal.

For SNe, the highest recall is obtained by the multi-scale model with 85.1 \%, followed closely by the full-size stamp scenario with 84.5 \%. Finally, the highest recall for variable stars is obtained by the cropped scenario (92.1 \%) and the multi-scale scenario (92 \%), both with very similar performance.  

\subsection{Impact on LSST}
The current specifications of LSST, as indicated in DMTN-102, say that the cutout stamps will be at least 30 x 30 pixels in size (900 floating point numbers). Considering a pixel size of $0.2\arcsec$, this translates to a minimum FoV of $6\arcsec\times6\arcsec$.

As the experiments with ZTF data show, a small FoV degrades the classification performance, especially for satellites and supernova. In consequence, we propose increasing the FoV of LSST stamps by using multi-scale stamps. One possibility that keeps the stamp payload close to the current LSST specification, but increases the FoV is to use 4 levels of 16 x 16 pixels each. The stamp sizes for this proposal are:

\begin{itemize}
    \item $16\times16$ pixels, with a $0.2\arcsec$ pixel width. FoV of $3.2\arcsec\times3.2\arcsec$.
    \item $16\times16$ pixels, with a $0.4\arcsec$ pixel width. FoV of $6.4\arcsec\times6.4\arcsec$.
    \item $16\times16$ pixels, with a $0.8\arcsec$ pixel width. FoV of $12.8\arcsec\times12.8\arcsec$.
    \item $16\times16$ pixels, with a $1.6\arcsec$ pixel width. FoV of $25.6\arcsec\times25.6\arcsec$.
\end{itemize}

This translates to 1024 floating point numbers. As one stamp is a high resolution version of the central part of the next stamp, an important part of the information is redundant between cutouts. In fact, 25 \% of the low resolution image can be computed from the next higher resolution stamp, which allows saving 18.75 \% of the space. The final size required for this 4-scale scheme is 832 floating point numbers, 7.56 \% smaller than the minimum cutout size described in the LSST specifications, but with a FoV 18.2 times larger in area.

This proposal for the LSST alert stamps is an educated guess based on the experiments performed with available data (ZTF). We plan to test the stamp classification with different combinations of scales and sizes when the first LSST alerts become available.

\section{Conclusions} \label{sec:conclusions}

We evaluated the use of a multi-scale approach for astronomical stamp classification, in the context of real-time classification of large streams of alerts. Using multi-scale image cutouts offers a good trade-off between a high resolution image of the source, a large FoV, and small file sizes. When comparing with other scenarios, such as ``cropped'', ``low resolution'', and ``full'', the multi-scale strategy offers the highest f1-score in the test set, although other models do have comparably high or higher metrics for certain classes.

The evaluation of the proposed model and strategy over unlabeled data shows a spatial distribution that is consistent with expectations for each class of object in the sky. The unlabeled stamps classified as satellites have a characteristic ``satellite glint'' or ``pearl necklace'' look.

Given the results presented here, we advocate that LSST adopts a multi-scale stamp strategy for the real-time alert stream. The current alert specifications of LSST have stamps with a very small FoV, which could negatively impact the ability of brokers to provide a high-quality fast transient classification. In fact, we expect the challenges of using small cropped images to be even greater in LSST versus what was tested here, considering the smaller angular size of LSST pixels in comparison with ZTF images. 

In the technical note DMTN-248 \citep{DMTN-248} of the Rubin Observatory, many options for the alert packets are discussed with the goal of maximizing scientific potential while minimizing alert packet size and required bandwidth, including the use of ``multi-resolution'' stamps. We believe that this work provides valuable new evidence in favor of using stamps with multiple scales, and therefore their adoption should be reconsidered.

Given that such a change in the LSST specifications would affect many different scientific uses, each one of the Science Collaborations of the Rubin Observatory should provide feedback about how a multi-scale stamp strategy would impact them before adopting any change. It is relevant to mention that the alert stream is not the only data product of LSST, and there are other ways to access the images for science cases that can afford longer delays between data acquisition and delivery to the scientific community.

\begin{acknowledgments}
    
The authors acknowledge support from the National Agency for Research and Development (ANID) grants: Millennium Science Initiative Program - ICN12\_009 (IRJ, FF, FEB, AMMA, GCV, GP, NA, PG, AM); BASAL Center of Mathematical Modelling Grant PAI AFB-170001 (IRJ, FF, AMMA, AA, NA); CATA-BASAL - ACE210002 (FEB) and FB210003 (FEB); FONDECYT Regular 1200710 (FF), 1190818 (FEB) and 1200495 (FEB); FB210005 (AMMA); and infrastructure funds QUIMAL140003 and QUIMAL190012; FONDECYT Iniciación 11191130 (GCV). A.B. acknowledges partial funding by the Deutsche Forschungsgemeinschaft Excellence Strategy - EXC 2094 - 390783311 and the ANID BASAL project FB210003. L.G. acknowledges financial support from the Spanish Ministerio de Ciencia e Innovaci\'on (MCIN), the Agencia Estatal de Investigaci\'on (AEI) 10.13039/501100011033, and the European Social Fund (ESF) ``Investing in your future'' under the 2019 Ram\'on y Cajal program RYC2019-027683-I and the PID2020-115253GA-I00 HOSTFLOWS project, from Centro Superior de Investigaciones Cient\'ificas (CSIC) under the PIE project 20215AT016, and the program Unidad de Excelencia Mar\'ia de Maeztu CEX2020-001058-M.

\end{acknowledgments}

\software{
    Astropy \citep{astropy:2013, astropy:2018, astropy:2022}, 
    Matplotlib \citep{Hunter:2007},
    NumPy \citep{harris2020array},
    Pandas \citep{reback2020pandas},
    Ray Tune \citep{liaw2018tune},
    scikit-learn \citep{scikit-learn},
    Tensorflow \citep{tensorflow2015-whitepaper}.
}

\appendix

\section{Classification models}
\label{sec:appendix:classification_models}
As a starting point, we took the Convolutional Network from \citet{carrasco2021alert} and made small adjustments to handle the different input sizes. As each stamp scale experiment has cutouts of different sizes and scales, it is not appropriate to use the same network for all the experiments. The architecture of the models used on each scenario are shown in Table \ref{table:architectures}.

\begin{table}[htbp]
\centering
\caption{Architectures used for the different scenarios. Input shape is indicated in the batch size, width, height, channels (b01c) standard. Layers are indicated in the same way as \citet{simonyan2014very}. For example, conv3-32 represents a convolutional layer with a 3 x 3 kernel size and 32 channels.
This table also contains the learning rates used to train each model. The number in the dropout layers is the fraction of units to be dropped.}
\label{table:architectures}
\begin{tabular}{cccc}
\hline
\multicolumn{4}{|c|}{ConvNet configurations}                                                                                                                      \\ \hline
\multicolumn{1}{|c|}{Full}              & \multicolumn{1}{c|}{Cropped}           & \multicolumn{1}{c|}{low\_res}          & \multicolumn{1}{c|}{Multi-scale}      \\ \hline
\multicolumn{4}{|c|}{Convolutional-pooling section}                                                                                                               \\ \hline
\multicolumn{4}{|c|}{Rotate and flip stamps}                                                                                                                      \\ \hline
\multicolumn{1}{|c|}{input 256x63x63x6} & \multicolumn{1}{c|}{input 256x16x16x6} & \multicolumn{1}{c|}{input 256x16x16x6} & \multicolumn{1}{c|}{input 256x8x8x24} \\ \hline
\multicolumn{1}{|c|}{conv3-37}          & \multicolumn{1}{c|}{conv3-112}         & \multicolumn{1}{c|}{conv3-63}          & \multicolumn{1}{c|}{conv3-76}         \\ \hline
\multicolumn{1}{|c|}{maxpool}           & \multicolumn{1}{c|}{conv3-27}          & \multicolumn{1}{c|}{conv3-31}          & \multicolumn{1}{c|}{conv3-18}         \\ \hline
\multicolumn{1}{|c|}{conv3-60}          & \multicolumn{1}{c|}{maxpool}           & \multicolumn{1}{c|}{maxpool}           & \multicolumn{1}{c|}{conv3-54}         \\ \hline
\multicolumn{1}{|c|}{maxpool}           & \multicolumn{1}{c|}{conv3-92}          & \multicolumn{1}{c|}{conv3-108}         & \multicolumn{1}{c|}{conv3-28}         \\ \hline
\multicolumn{1}{|c|}{conv3-29}          & \multicolumn{1}{c|}{conv3-121}         & \multicolumn{1}{c|}{conv3-14}          & \multicolumn{1}{c|}{}                 \\ \hline
\multicolumn{1}{|c|}{maxpool}           & \multicolumn{1}{c|}{}                  & \multicolumn{1}{c|}{}                  & \multicolumn{1}{c|}{}                 \\ \hline
\multicolumn{1}{|c|}{conv3-26}          & \multicolumn{1}{c|}{}                  & \multicolumn{1}{c|}{}                  & \multicolumn{1}{c|}{}                 \\ \hline
\multicolumn{4}{|c|}{Average rotated and flipped feature maps}                                                                                                    \\ \hline
\multicolumn{4}{|c|}{Fully-connected section}                                                                                                                     \\ \hline
\multicolumn{1}{|c|}{dropout-0.71726}   & \multicolumn{1}{c|}{dropout-0.866799}  & \multicolumn{1}{c|}{dropout-0.73065}   & \multicolumn{1}{c|}{dropout-0.8478}   \\ \hline
\multicolumn{1}{|c|}{FC-31}             & \multicolumn{1}{c|}{FC-133}            & \multicolumn{1}{c|}{FC-22}             & \multicolumn{1}{c|}{FC-69}            \\ \hline
\multicolumn{1}{|c|}{FC-6}              & \multicolumn{1}{c|}{FC-6}              & \multicolumn{1}{c|}{FC-6}              & \multicolumn{1}{c|}{FC-6}             \\ \hline
\multicolumn{4}{|c|}{Model output}                                                                                                                                \\ \hline
\multicolumn{1}{l}{}                    & \multicolumn{1}{l}{}                   & \multicolumn{1}{l}{}                   & \multicolumn{1}{l}{}                  \\ \hline
\multicolumn{4}{|c|}{Learning rate}                                                                                                                               \\ \hline
\multicolumn{1}{|c|}{$7.589\times 10^{-4}$}          & \multicolumn{1}{c|}{$3.4627\times 10^{-4}$}         & \multicolumn{1}{c|}{$1.125\times 10^{-3}$}          & \multicolumn{1}{c|}{$7.445\times 10^{-4}$}         \\ \hline
\end{tabular}
\end{table}

In all the models, we rotated the inputs by $0^{\circ}, 90^{\circ}, 180^{\circ} \text{ and } 270^{\circ}$, and also used their up-down flipped versions, giving a total of 8 variations for each input stamp. Each one of the 8 variations is processed by the same convolutional and pooling layers, and after that the 8 representations are averaged and passed to the final dense layers.

After the convolutions, pooling, and averaging of the 8 variations, a dropout operation is done and the position metadata is concatenated to the representation coming from the stamps. This representation with stamp and position information passes through two dense (fully-connected) layers.

The architectures mainly differ in the number of pooling operations, in which we took the input size into consideration. The number of pooling layers chosen leads to 8x8 feature maps at the end of the convolutional - pooling stage in all four scenarios.
    
The batch size was fixed at 256 samples. Each sample on the batch is chosen randomly from the different classes with equal probability, so the batches are balanced on average. The adopted loss function was cross-entropy and the optimizer was ADAM. The stopping criteria looked at the validation F1-score every 500 iterations and stopped the training if that metric did not improve in 5 evaluations with respect to the best F1-score recorded so far.

In order to give each scenario a fair chance to perform the best, we ran a hyperparameter search using Ray Tune \citep{liaw2018tune} for each of them. The hyperparameters explored were the number of convolutional filters on each layer, the size of the first dense layer, the learning rate, the dropout rate, the kernel size of the first convolution (between 3x3 and 5x5 pixels) and to use or not batch normalization after the source position is given to the network. The hyperparameter search was done with HyperOpt \citep{bergstra_making_2013} and the experiment scheduler used was the Asynchronous HyperBand Scheduler \citep{li_system_2020}, both available in Ray Tune.

When doing the hyperparameter search we checked that the parameters of the best model were not too close to the boundary of the search space. If that were the case it might indicate that increasing the search range could allow the discovery of even better solutions. In the case that the best model found by the hyperparameter search algorithm was very close to the search range (less than 10 \% of the search range), the hyperparameter search was repeated with an expanded range.

\section{Classification metrics}
\label{sec:classification_metrics}

In this paper we compute 3 classification metrics: precision, recall and f1-score. All these metrics were calculated as macro metrics. To define these metrics we must introduce some concepts:

\begin{itemize}
    \item $TP_{i}$: True positives of class $i$. Amount of objects from class $i$ that are classified as $i$.

    \item $FP_{i}$: False positives of class $i$. Amount of objects classified as $i$ but that do not truly belong to class $i$.

    \item $FN_{i}$: False negatives of class $i$. Amount of objects from class $i$ that were not classified as $i$.

    \item $\text{Precision}_{i} = TP_{i}/(TP_{i} + FP_{i})$.

    \item $\text{Recall}_{i} = TP_{i}/(TP_{i} + FN_{i})$.

    \item $\text{F1-score}_{i} = 2\cdot \text{Precision}_{i} \cdot \text{Recall}_{i} / (\text{Precision}_{i} + \text{Recall}_{i})$.
\end{itemize}

In a classification problem with N classes, we can define the macro metrics as follows:

\begin{equation}
    \text{Precision}_{\text{macro}} = \frac{1}{N}\sum_{i \in \text{classes}} \text{Precision}_{i}
\end{equation}

\begin{equation}
    \text{Recall}_{\text{macro}} = \frac{1}{N}\sum_{i \in \text{classes}} \text{Recall}_{i}
\end{equation}

\begin{equation}
    \text{F1-score}_{\text{macro}} = \frac{1}{N}\sum_{i \in \text{classes}} \text{F1-score}_{i}
\end{equation}

All of these metrics take values between zero and one, although in the paper they are presented as ``percentages'' (between zero and one hundred).

\section{Spatial distribution in the horizontal coordinate system}
\label{sec:spatial_distribution_altaz}

\begin{figure}[htbp]
    \centering
    \includegraphics[width=\textwidth]{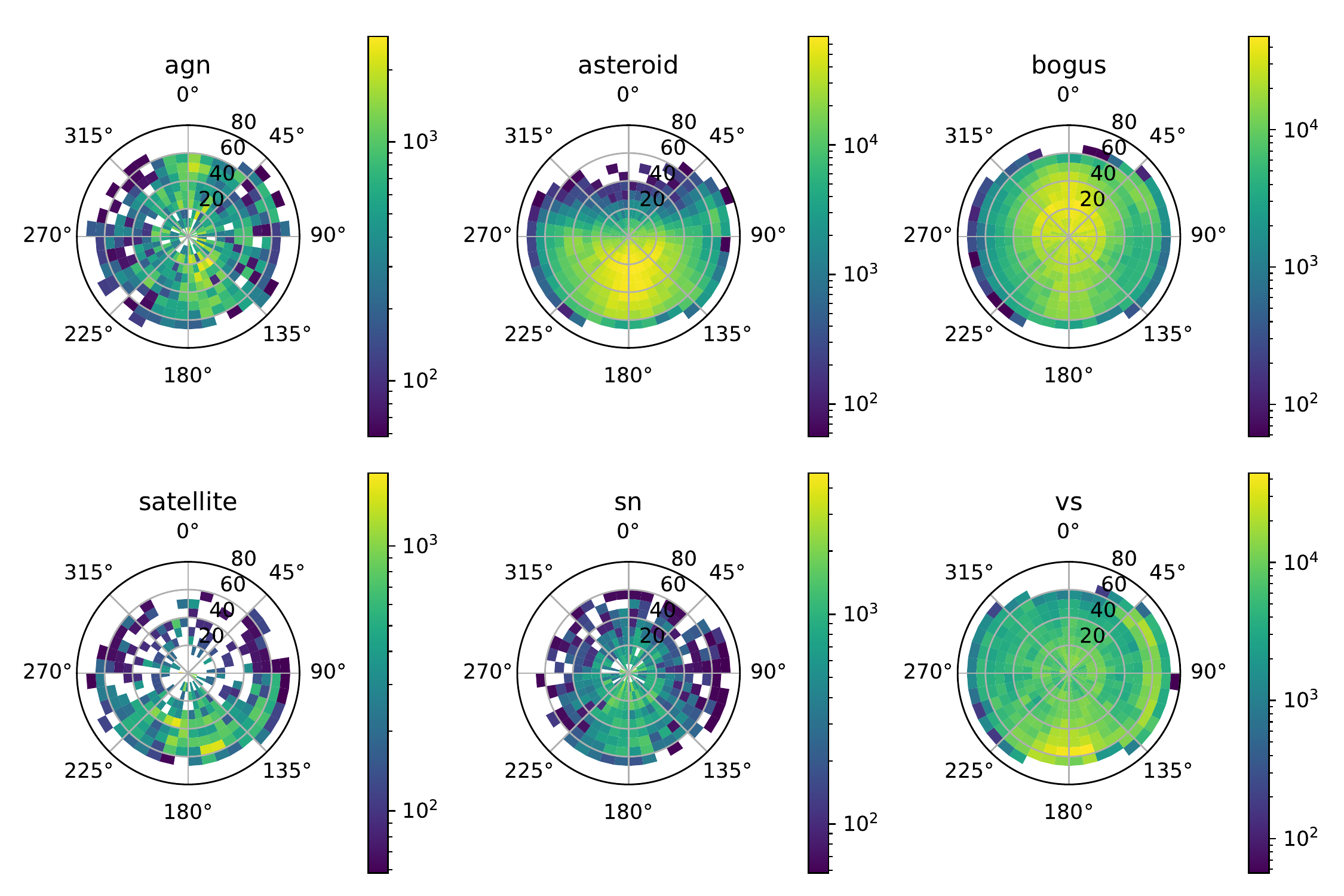}
    \caption{Altitude - Azimuth spatial distribution of the sources shown in figure \ref{fig:unlabeled_spatial_distribution}. The numbers in the color bar represent the object density, i.e. the number of sources per squared radian (steradian).}
    \label{fig:unlabeled_spatial_distribution_altaz}
\end{figure}

Figure \ref{fig:unlabeled_spatial_distribution_altaz} shows the same sources as figure \ref{fig:unlabeled_spatial_distribution} but in the local Altitude - Azimuth coordinate system from Palomar Observatory, where the ZTF survey is done. The plots were made by taking the complementary angle of the altitude, so that 0 radius (center of the plot) represents the zenith position. The figure shows that most objects classified as asteroids are located to the south, which is consistent with the fact that asteroids lie on the ecliptic plane and are observed from a location which is north of the parallel $23^{\circ}\; 26'$ N. Most objects classified as satellites are also located to the south direction. This is expected for geostationary satellites and for satellites orbiting close to the Equatorial plane, given the location of Palomar Observatory. AGN and SN have relatively random locations. Variable stars preferentially lie to the south, as the greater portion of the Galactic Plane lies south of Palomar as well.

\section{Large images of satellite candidates}
\label{sec:large_fov_satellite}

\begin{figure}[htbp]
    \gridline{
    \fig{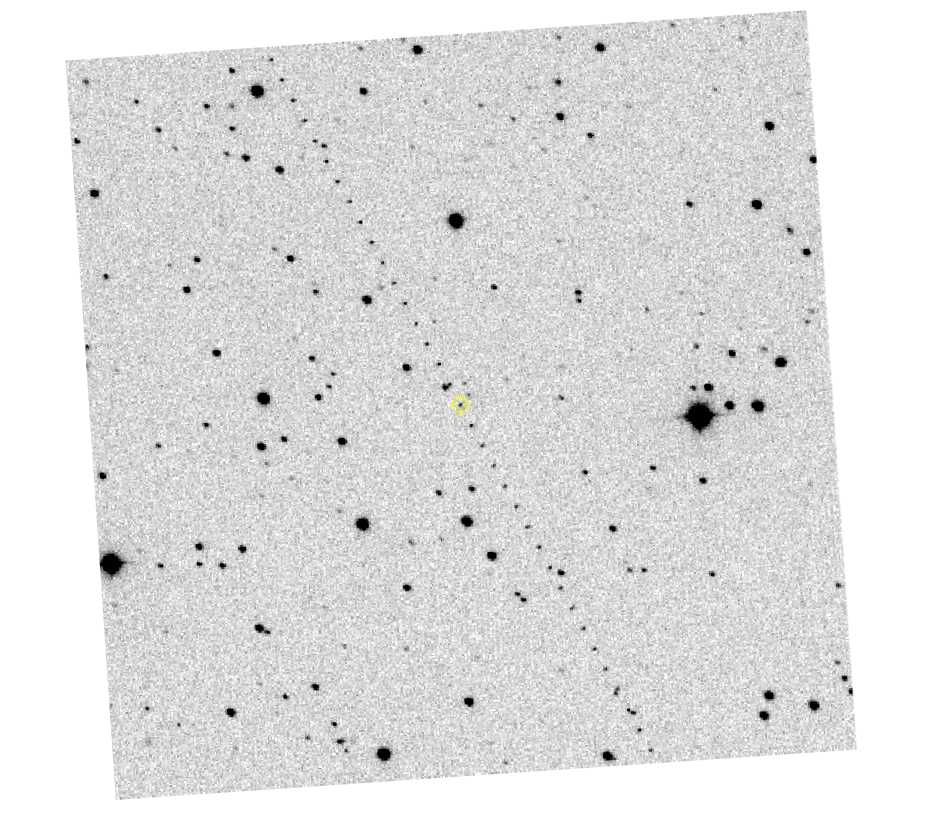}{0.45\textwidth}{(a) ZTF21abklvrx}
    \fig{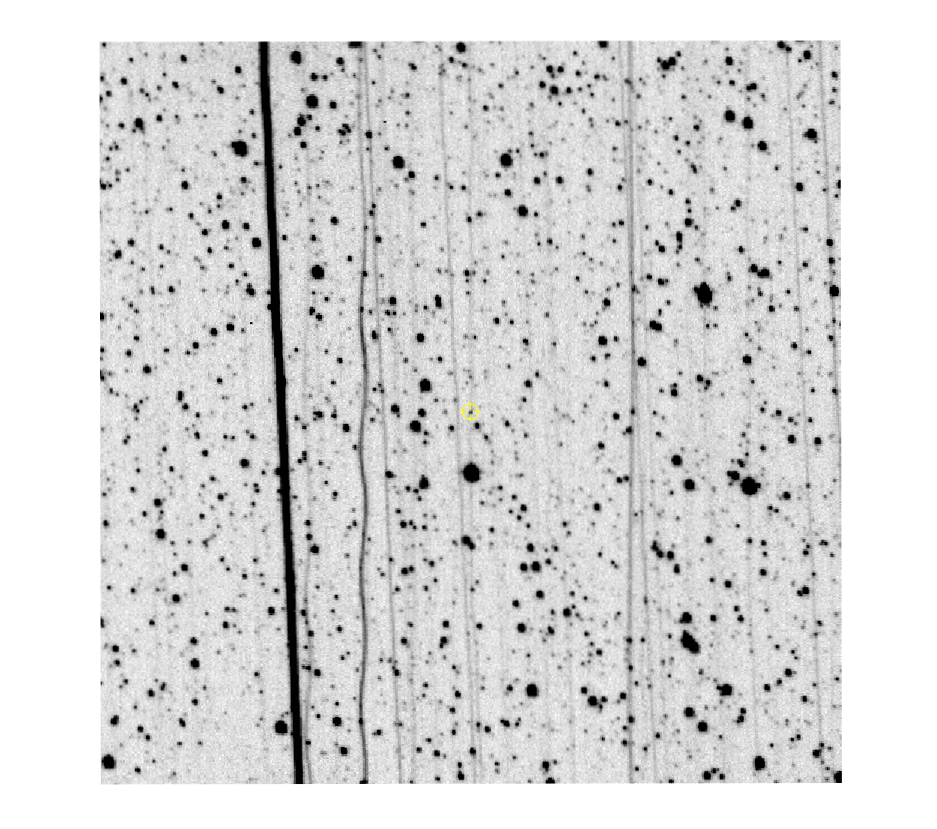}{0.45\textwidth}{(b) ZTF20achpliw}}
    \caption{$500''\times 500''$ images for ZTF alerts ZTF21abklvrx and ZTF20achpliw.}
    \label{fig:large_fov_satellites}
\end{figure}

We select two of the satellite candidates from Figure \ref{fig:satellite_candidates} and examine their respective science images in the NASA/IPAC Infrared Science Archive \citep{https://doi.org/10.26131/irsa539}. Figure \ref{fig:large_fov_satellites} shows both images. ZTF21abklvrx has a clear ``pearl necklace'' look, which suggests that this alert corresponds to a rotating satellite. When looking at the full image for ZTF20achpliw we see many vertical lines across the observation. According to the ZTF team (private communication), this event is probably caused by a readout and clock error in the instrument. 

\bibliography{sample631}{}
\bibliographystyle{aasjournal}

\end{document}